\documentclass[usenatbib]{mn2e}
\usepackage{graphicx}
\usepackage{url}
\usepackage{lmodern}    
\usepackage[T1]{fontenc}

\newcommand {\kms}{km\,s$^{-1}$}
\newcommand {\andxv}{And~XV}
\newcommand {\andxvi}{And~XVI}

\def\ltsima{$\; \buildrel < \over \sim \;$}
\def\lta{\lower.5ex\hbox{\ltsima}}
\def\gtsima{$\; \buildrel > \over \sim \;$}
\def\simgt{\lower.5ex\hbox{\gtsima}}
\def\kms{{\rm\,km\,s^{-1}}}

\def\msun{{\rm\,M_\odot}}

\def\AA{$\; \buildrel \circ \over {\rm A}$}

\def\s{\ifmmode \widetilde \else \~\fi}
\def\={\overline}

\def\spose#1{\hbox to 0pt{#1\hss}}

\def\lta{\mathrel{\spose{\lower 3pt\hbox{$\mathchar"218$}}
     \raise 2.0pt\hbox{$\mathchar"13C$}}}
\def\gta{\mathrel{\spose{\lower 3pt\hbox{$\mathchar"218$}}
     \raise 2.0pt\hbox{$\mathchar"13E$}}}
\def\Dt{\spose{\raise 1.5ex\hbox{\hskip3pt$\mathchar"201$}}}    
\def\dt{\spose{\raise 1.0ex\hbox{\hskip2pt$\mathchar"201$}}}    

\def\dotsfill{\leaders\hbox to 1em{\hss.\hss}\hfill}

\def\ec4{EC4}

\def\andxv{And\,XV}
\def\andxvi{And\,XVI}

\loadboldmathitalic

\title[]{A Keck/DEIMOS spectroscopic survey of the faint M\,31 satellites \andxv\ and \andxvi}
\author[Bruno\ Letarte et al.] 
{B. Letarte$^1$, S. C. Chapman$^2$, M. Collins$^2$, R. A. Ibata$^3$, M. J. Irwin$^2$,  
\newauthor
A. M. N. Ferguson$^4$,
G. F. Lewis$^5$, N. Martin$^6$, A. McConnachie$^7$, N. Tanvir$^8$
\\
$^{1}$ California Institute of Technology, 1200 E. California Blvd, MC 105-24, Pasadena, CA 91125 USA\\
$^{2}$ Institute of Astronomy, Madingley Road, Cambridge, CB3 0HA, U.K.\\
$^{3}$ Observatoire de Strasbourg, 11, rue de l'Universit\'e, F-67000, Strasbourg, France\\
$^4$ 
Institute for Astronomy, University of Edinburgh, Royal Observatory, Blackford Hil l, Edinburgh, UK EH9 3HJ\\ 
$^5$ 
Institute of Astronomy, School of Physics, A29, University of Sydney, NSW 2006, Australia\\ 
$^6$ 
Max-Planck-Institut fur Astronomie, Konigstuhl 17, D-69117 Heidelberg, Germany\\ 
$^7$ 
Department of Physics and Astronomy, University of Victoria, Victoria, B.C., V8P 1A1, Canada\\ 
$^8$ 
Department of Physics \& Astronomy, University of Leicester, Leicester, LE17RH, UK\\
}

\date{\today}
\begin{document} 

\maketitle 
\begin{abstract} 
We present the results of a spectroscopic survey of the recently discovered M31 satel- 
lites \andxv\ and \andxvi, lying at projected distances from the centre of M31 of 93 
and 130~kpc respectively. These satellites lie to the South of M31, in regions of the 
stellar halo which wide field imaging has revealed as relative {\it voids} (compared to the $\sim$degree-scale coherent stream-like structures). 
Using the DEep Imaging Multi-Object Spectrograph mounted on the Keck II 
telescope, we have defined probable members of these satellites, for which we derive 
radial velocities as precise as $\sim6~\kms$\
down to $i\sim$21.5. 
While the distance to \andxvi\ remains the same as previously reported 
(525$\pm$50 kpc), we have demonstrated that the brightest three stars previously used 
to define the tip of the red giant branch (TRGB) in \andxv\ are in fact Galactic, and \andxv\ is 
actually likely to be much more distant at 770$\pm$70 kpc (compared to the previous 630~kpc), increasing the luminosity from M$_{\rm V}\approx-9.4$ to M$_{\rm V}\approx-9.8$. 
The \andxv\ velocity dispersion is resolved 
with v$_r$ =-339$^{+7}_{-6} \kms$
and $\sigma_v = 11^{+7}_{-5} \kms$. 
The \andxvi\ dispersion is not quite resolved at $1\sigma$ with v$_r$ =-385$^{+5}_{-6} \kms$
and $\sigma=0^{+10}_{-indef} \kms$.
Using the photometry of the confirmed member stars, we find metallicities of And\,XV (median [Fe/H]$=-1.58$, interquartile range $\pm$0.08), and And\,XVI (median [Fe/H]$=-2.23$, interquartile range $\pm$0.12). 
Stacking the spectra of the member stars, we find spectroscopic [Fe/H]$=-1.8$ (-2.1) for \andxv\ (\andxvi), 
with a uncertainty of  $\sim 0.2$~dex in both cases.
Our measurements of \andxv\ reasonably resolve its mass ($\sim10^8$ M$_\odot$) and 
suggest a polar orbit, while the velocity of \andxvi\ suggests it is approaching the 
M31 escape velocity given its large M31-centric distance.
\end{abstract}
 

\section{Introduction}

In recent years, systematic searches have been performed in earnest within the
Local Group, both photometrically and spectroscopically, in order to fully
catalogue its population of satellite galaxies. There are several motivations
behind these endeavours, with one of the more hotly discussed being that of the
so-called ``missing satellites problem'' \citep{moore99,klypin99}, which is an
observed discrepancy of 1-2 orders of magnitude between the number of satellite
galaxies produced within cosmological simulations and the number we see
orbiting the Galaxy and M31 today. And whilst these surveys have approximately
doubled the known satellite populations of both galaxies within recent years
\citep{zucker04,willman05, chapman05,belokurov06a,martin06,zucker06a,zucker06b,belokurov07a,majewski07, belokurov07a,irwin08} we are still far from reconciling the
observations with simulations. Several theories have been put forward in order
to address this issue. Survey completeness is a widely debated topic (see e.g.
\citealt{tollerud08,simon07}), particularly with respect to the MW, where
observational efforts are hampered by obscuration from the disk and the bulge.
Future all-sky surveys may find a wealth of
ultra-faint satellites that would allieviate this gap between observation and
theory. Another proposed solution is that the satellite galaxies within the
local group inhabit increasingly dark matter dominated halos as they become
fainter \citep{bullock00,stoehr02}, so that even the faintest of dwarf galaxies
would reside within massive dark matter halos. 

Penarrubia et al. (2006, 2008a, 2008b) demonstrate that within
the LCDM working frame, dwarf spheroidal galaxies (dSphs) are well embedded within the dark
matter halos that surround them. For the brightest dwarf galaxies in the
MW they find that $0.01< R_c/r_s <0.1$, where $R_c$ and $r_s$ are the
stellar core radius  and the NFW scale radius, respectively. That
implies that these systems are fairly resilient to tidal interactions,
so that they can lose a large fraction of their initial dark matter mass before the
stellar component starts being disrupted. 

An observational test of these 
suggestions is then to measure the stellar light and dynamical mass of faint
dwarf galaxies in the Local Group. With spectroscopic information, one can
measure the central velocity dispersion of the satellite, which has been shown
to be a good indicator of the instantaneous mass of a dwarf galaxy within the
radii of the luminous tracers \citep[e.g.][]{oh95,piatek95}, even if the dwarf
is not in virial equilibrium or perfectly spherical. Several approaches have
been taken to derive the total mass of the system from the central velocity
dispersion \citep[e.g.][]{illingworth76,richstone86,gilmore07}. Recent detailed
modelling of dwarf galaxy velocity dispersions suggests a lower dynamical mass
limit of $\sim10^7\msun$ (Strigari et al.\ 2008).

Spectroscopic data on the dSphs of the local group is also
desirable in order to give us a better understanding of the nature of these
tiny galaxies. The discovery and study of the faintest members of this
population has shown that at the low luminosity end of the spectrum, these
satellites do not behave as expected from the study of their brighter
counterparts. In particular, they begin to deviate significantly from
established trends between mass and metallicity, and M/L vs light.
In the brighter MW dwarfs,
the lack of very metal poor stars \citep{helmi06}, the stellar
populations \citep{unavane96} 
and differing
abundance patterns from the MW halo \citep{Shetrone01} reinforce the
point that the surviving, isolated dwarfs have different environmental
conditions than those which phase mixed into the MW stellar halo.
However, the ultra-faint MW satellites do show evidence of very metal poor stars
\citep{kirby08}, possibly suggesting these dwarfs are remnants of the halo building blocks.

By aquiring kinematic data for the Local Group satellites, we
can isolate the likely dwarf member stars from the Colour Magnitude Diagrams
(CMD), perform independent checks on the metallicities derived photometrically,
and also estimate their total masses so we can place them more fully in the
context of the dSph population, and determine if there truly is a breakdown in
these relations at low luminosities, and if so, at what point this divergence
begins to take effect.

Another motivation behind obtaining kinematic data for the satellite
populations of the MW and M31 is to achieve a greater understanding of the
dynamics of the Local Group as a whole. The mass contained within the inner few
10's kpc of both is relatively well constrained from optical and HI data, as
well as globular cluster and planetary nebulae tracer populations (see e.g.
\citealt{kochanek96,carignan06}), but probing the outer halo at large
radii proves to be more of a challenge. An ideal method for measuring the {\it
total} masses of the MW and M31 is to use their satellite galaxies as direct
tracers of their potentials. As they are located at large distances from their
hosts ($\sim$few 100's kpc) they probe the full extent of their mass
distributions. Such a method has been used previously in both the MW and M31
\citep{evans00,wilkinson99,gottesman02}, however the associated
errors on their measurements are huge (of order $\sim$twice the measurements
themselves) due to the low number of satellite galaxies with reliable kinematic
and distance data available to them. In the years since these results were
published, more than 20 satellite companions of the MW and M31 have been
discovered, many of which have also been studied spectroscopically. If
kinematic data on all the satellites within the Local Group can be obtained, we
will be able to better constrain the masses of these two ``sister'' galaxies. 

With these goals in mind, it remains crucial to obtain radial velocities for
newly discovered dSphs in order to understand their mass distribution and
orbital properties. As a step towards this end, we have used the Keck II DEep
Imaging Multi-Object Spectrograph to derive radial velocities and metallicities
of stars within two new satellites discovered in \citet{ibata07}, \andxv\ and
\andxvi.

\section{Observations}
Multi-object Keck observations with DEIMOS \citep{faber} for \andxv\
and \andxvi\ were made on 2007 Oct 8--9, in variable
conditions (with 0.6--1.0$''$ seeing and patchy cirrus).  
To obtain improved S/N in a reasonable exposure time for 
the faint ($i = 20.5-22.5$) RGB stars targeted, we employed 
the lower resolution 600 line/mm grating. For the Ca{\sc II} triplet (CaT) lines, 
which are reasonably resolved at this lower (R$\sim$3000) resolution, the trade-off is a strong function of the location of 
the spectral features of interest relative to the strong night 
sky OH recombination lines. While we detect continuum at 
S/N$>$3 per \AA\  over the CaT regions 
for faint ($i\sim22$) stars, the velocity errors can suffer larger systematic errors when they lie close to OH lines, and the velocity accuracy can vary considerably with continuum S/N.
Our chosen instrumental setting covered the observed 
wavelength range from $0.56-0.98 \mu$m. Exposure time 
was 60 min on each dSph, split into 20-min integrations. Data reduction 
followed standard techniques using the DEIMOS-DEEP2 
pipeline (Faber et al. 2003), debiassing, flat-fielding, extracting, wavelength-calibrating and sky-subtracting the spectra. 

The radial velocities of the stars were then measured by 
fitting the peak of the cross-correlation function derived 
using a template spectrum consisting of delta functions to the instrumental resolution (3\AA) at the wavelengths of the CaT absorption lines. This procedure also provides an estimate of the radial velocity accuracy obtained 
for each measurement. The velocity uncertainties typically range from 
$6$ to $25 \kms$.  
Target stars were assigned from a broad (1 mag) box 
around the general outline of the RGB of the dwarf. 
A narrower box of 0.3 mag around the RGB was then constructed with 
higher priorities. In both of these selection boxes, the priority was scaled as a function of $i$-mag
so that brighter stars 
had more chances of being observed. The remainder of the 
mask space was filled with target stars from a broad region 
encompassing RGB stars in M31 over all possible metallicities and distances. The And\,XV, And\,XVI masks had 143, 131 target 
stars respectively. The small number of candidate members 
compared to the total amount of stars observed is simply 
due to the fact that the footprint of DEIMOS is much wider 
than the small angular size of the dwarf galaxies, therefore, 
there are many more stars that are not part of the main dSph targets. 

\section{Results}

The spatial distributions of the stars in the regions of 
\andxv, \andxvi\ are shown in Figure 1. Candidate member stars from the DEIMOS spectroscopy are highlighted (listed in Tables 1 \& 2).  
Member stars are selected by having a velocity which lies within $2\sigma$ of the median of the obvious kinematic clumps of stars in each dwarf field.
Note that for \andxv, the member stars are highlighted differently for {\it robust} and {\it tentative} members, defined below in terms of the strength of the cross-correlation peak in the velocity determination. 
CFHT-MegaCam colour-magnitude 
diagrams (CMDs) of stars within a three arcminute radius of 
\andxv\ and \andxvi\ are shown in Figure 2, again with 
likely member stars highlighted. 
The red giant branches of 
both dwarfs are clearly visible, while the horizontal branches are reasonably detected. 
In the case of \andxv, 
three stars previously assumed to lie near the TRGB have 
been shown by their velocities and NaI doublet equivalent width (EW) to be foreground Galactic dwarf stars. We then revise the distance estimate 
of \andxv\ from the 630$\pm$60~kpc previously reported (Ibata et al. 
2007), estimating the true tip of the RGB is $\sim$0.5 mag fainter 
than the $i=20.4$ previously assumed. Adopting MTRGB = 
-4.04 $\pm$ 0.12 from Bellazzini, Ferraro \& Pancino (2001) for 
the absolute $I$-band magnitude of the RGB tip, and convert 
into the Landolt system using the colour equations in Ibata et al.\ (2007)
and those given by McConnachie et al.\ (2004); this yields 
a distance modulus of $m - M = 24.4\pm0.2$ or a distance 
of 770$\pm$70~kpc (versus the previous 630$\pm$60~kpc). 
This revised distance agrees with that obtained by considering the magnitude, $g \sim 25.3$, of the 
Horizontal Branch (Fig.~2).
The change in distance increases the luminosity
from M$_{\rm V}\approx -9.4$ to M$_{\rm V}\approx -9.8$. 
In \andxvi, as all stars near the TRGB were confirmed 
spectroscopically as  members of the dwarf spheroidal galaxy, the Ibata et al.\ (2007) 
distance estimate of 525 $\pm$ 50 kpc remains unchanged.  
The  horizontal branch for \andxvi\ (Fig.~2) at $g\sim24.5$ (0.8~mag difference from \andxvi) agrees well with our proposed 0.5~mag difference in the TRGB between the two dwarf spheroidals.
In Fig.~2, we also overlay 13~Gyr old Padova isochrones with solar-scaled chemical compositions  (Girardi et al.\ 2004), at the TRGB distance and median metallicity obtained from member stars (see below), providing a reasonable fit in both cases.
The use of an old (13 Gyr) isochrone can be justified by the predominantly old nature of the majority of the recently detected M31 satellites.   

The distribution of stars in And\,XV and And\,XVI is shown in Fig.~3 as a function of their radial velocity. 
The stars are then shown as 
a function of radius from the centers of the dwarfs with half-light radii indicated. 
Photometric [Fe/H] is estimated by comparison to Padova isochrones (Girardi et al.\ 2004) corrected for extinction \citep{schlegel} and shifted to the revised distances of the dSphs. 
These [Fe/H] values are shown in Fig.~3 as a function of their radial velocity, 
revealing the tight range in 
metallicities of both And\,XV (median [Fe/H]=-1.58, interquartile range $\pm$0.08), And\,XVI (median [Fe/H]=-2.23, interquartile range $\pm$0.12). For And\,XV the  median [Fe/H] and its dispersion are identical using either the robust spectra or including the tentative spectra as well, lending further evidence to these stars likely being \andxv\ members.

One-dimensional spectra of member stars in 
\andxv\ and \andxvi\ are shown in Figs.\ 4 \& 5 respectively. \andxv\ spectra 
are separated into robust members with 
cross-correlation peak $> 0.2$, and more tentative member stars 
with 
cross-correlation peak $< 0.2$, but still lying 
on the well defined RGB of \andxv. The inverse variance 
weighted, summed spectrum is shown in the bottom offset 
panel for each dSph. All candidate member spectra for \andxvi\
are considered to be robust by the above criteria. While 
not shown in the spectra, the Na{\sc I} doublet is undetected 
significantly in the individual stars, and also in the stacked 
spectra. The Na{\sc I} equivalent width is sensitive to the surface gravity, and thus 
is a good discriminant of Galactic foreground dwarf stars \citep{schiavon97},
although at these large negative velocities, the probability is 
very low that any identified star in our CMD selection box 
would be Galactic. Stars with velocities $<150 \kms$, consistent with Galactic stars typically have well detected Na{\sc I} doublet lines in our DEIMOS spectra.
The spectroscopic [Fe/H] has very large errors in most of the individual spectra, however a reasonable comparison can be made between the [Fe/H] derived from the  stacked spectrum and the median photometric [Fe/H]. We find  [Fe/H]$=-1.8$ (-2.1) for \andxv\ (\andxvi) by measuring the EW of the Ca{\sc II} triplet lines as in Chapman et al.\ (2005), on the \citet{carretta} scale. A large uncertainty, measured from the summed sky spectrum, of  $\sim 0.2$~dex is due primarily to sky-subtraction residuals making it  difficult to define the continuum level  of  the  spectrum. 
Koch et al.\ (2008) have also demonstrated that high quality Keck/DEIMOS spectra of stars in the M31 halo are amenable to further chemical analysis, showing a range of species (mostly Fe{\sc I} and Ti{\sc I} lines) which become weaker for the more metal-poor stars.
For \andxv\ and \andxvi, even the stacked spectra do not detect significant absorption at the Ti{\sc I} lines
(8378, 8426, and 8435\AA). Our stacked spectra do not have the requisite S/N to detect the anticipated EW based on stars of similar metallicity from Koch et al.\ (2008), although roughly tripling the S/N of the spectra  would be more than sufficient ($\sim4\times$ integration time in typical Mauna Kea weather conditions, compared to the rather poor conditions under which the present data were taken). 

For our samples of member stars with large and variable velocity errors (typically 6--25 $\kms$), the Maximum Likelihood approach (e.g., Martin et al. 2007) provides a method to assess the true underlying velocity distribution of the dSphs. Figure~6 shows the results of our analysis of \andxv\ and \andxvi\ plotting 
the one dimensional distributions 1, 2, 3$\times \sigma$ values. 
The \andxv\ dispersion is resolved 
with v$_r$ =-339$^{+7}_{-6} \kms$
and $\sigma_v = 11^{+7}_{-5} \kms$, where little difference 
is found from using only the best seven member stars (from 
Fig.~4) versus using all 13 candidate members (including the 
six more tentative velocities measurements). 
The \andxvi\ dispersion is not quite resolved at $1\sigma$ with v$_r$ =-385$^{+5}_{-6} \kms$
and $\sigma=0^{+10}_{-indef} \kms$.

\section{Discussion and Conclusions}

We have been able to further constrain the properties of 
the dSphs And\,XV, And\,XVI by measuring velocities and 
metallicities and assigning probable member stars to 
each galaxy. While this represents a significant addition to our knowledge about these dSphs compared to the photometric discovery and characterization (Ibata et al.\ 2007), 
our spectroscopic measurements are generally not precise 
enough to provide robust measures of the velocity dispersions. 
\andxv\ has a most likely dispersion $\sim10 \kms$ (while \andxvi\ is more poorly constrained but has a $1\sigma$ upper limit of $10~\kms$) which would
translate  to a $\sim10^8$~M$_\odot$ halo mass using the method of \citet{richstone86} as has been
done for other M31 dSphs with better constrained velocity dispersions (e.g., Chapman et al.\ 2005, 
Majewski et al.\ 2007). This is similar to that proposed by Gilmore et al.\ (2007) for a limiting dark matter halo mass in the smallest galaxies, however we cannot confidently rule out that the dispersions are 
smaller than our measurements suggest. Longer Keck/DEIMOS exposures under good conditions could significantly improve these results. 
The relatively well populated RGBs of both dSphs suggests 
that enough member stars could be obtained to begin to 
measure the velocity dispersion profile, a much more reliable constraint on the dark matter halo mass (Gilmore et al.\ 2007). 

As mentioned, both dSphs lie in relative voids (free of degree-scale
substructure) in the M31 halo maps of Ibata et al.\ (2007). In 
our spectroscopic samples there are relatively few remaining 
M31 halo stars once the likely And\,XV, And\,XVI member 
stars and Galactic foreground are removed (3 and 6 candidate halo stars respectively, and some of these are likely to be Galactic by their proximity to the low-velocity regime of Milky Way stars). This is consistent with the extrapolated M31 halo profle (Ibata et al.\ 2007) out to 93 kpc 
and 130 kpc respectively. 
However, this suggests that significant spectroscopic efforts would be required to properly characterize 
the (substructure-free) halo in the 100-150 kpc regime. Nonetheless, the average spectroscopic [Fe/H]=-1.5 of these 9 confirmed M31 
halo stars is consistent with a lack of metallicity grandient in 
the underlying metal poor ([Fe/H]=-1.4) halo found within 
the inner 70~kpc of M31 (Chapman et al.\ 2006). Spectroscopic studies of RGB stars along the minor axis of M31 have been prone to inadvertently sampling stream-like substructure (Gilbert et al.\ 2006; Koch et al.\ 2008; Chapman et al.\ 2008), but they also do find a range of metallicities in the 100-150~kpc regime consistent with the [Fe/H] in these dSph fields.

We have described before in Chapman et al.\ (2008) 
the expected probabilities of halo star contamination to fields at these
distances. Here we run a simple Monte Carlo simulation of the expected effect of the
contaminant on the observed $\sigma_v$.
We take two distributions, for the halo (120km/s $\sigma_v$ at ~100kpc
projected) and for each dSph as measured here.
We then draw randomly from these distributions, asking how often halo
stars land in some window indistinguishable from the dSph.
We then remeasure the dwarf $\sigma_v$ whenever star(s) from the halo are
present.
We find a halo star lands in the \andxv\ window 25\% of the time, and in
\andxvi\ 9\% of the time.
In 100 samples with a contaminating halo star(s) lying in the dSph, the
velocity dispersion was measured to range 11-12 km/s (\andxv, assumed 
$\sigma_v$=11km/s) and 8-9 km/s (\andxvi, assumed $\sigma_v$=8 km/s).
Thus the affect of halo stars lying in the velocity window is small    
relative to the errors in estimating $\sigma_v$.

We then turn to the orbital properties of these dSphs. 
And\,XV lies near the minor axis of M31, at about the same distance and heliocentric velocity as M31 
(785~kpc and v$_r\sim$-300 $\kms$). As such, its orbit must be close to polar with a large implied 
tangential velocity component. By contrast, at the large 
M31-centric distance of And\,XVI ($\sim$270 kpc), its velocity of 
$\sim-400 \kms$ pushes it towards the escape velocity of M31 (assuming the mass modeling of Geehan et al.\ 2006; see Chapman et al.\ 2007 Fig.~4 for a depiction of the M31 dwarf galaxies in this context).
This is comparable to the orbital properties other recently discovered dSphs (And\,XII: Chapman et al.\ 2007; 
And\,XIV:  Majewski et al.\ 2007). In particular, And\,XII is traveling near the escape velocity of the entire Local Group. Finding so many satellites near the escape velocity of M31 may suggest that the total mass of M31 has been somewhat underestimated to date.
 
\andxvi\ appears to sit spatially within a preferred distribution of satellites of M31 (Koch et al.\ 2006; McConnachie \& Irwin 2006), 
falling towards us at $\sim100 \kms$ (with an unknown tangential component).
However,  \andxv\ sits somewhat outside of this 
distribution (in both position and likely orbit) and may represent one of the growing number of discovered M31 satellites which hint at a more complicated environment than was 
apparent with the first generation of discovered satellites.

\begin{figure}
\centering
\includegraphics[width=0.5\textwidth, angle=0]{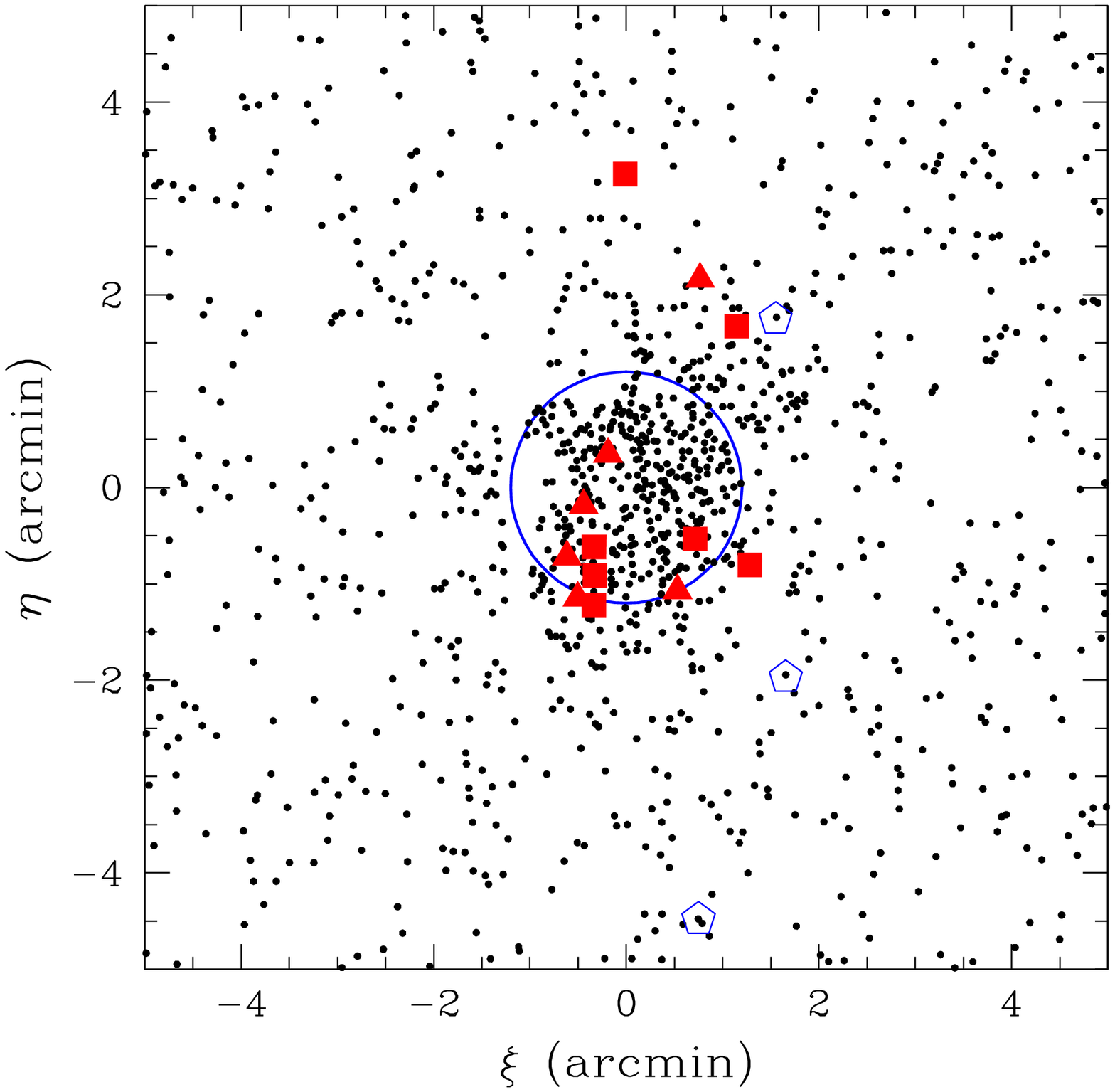}
\includegraphics[width=0.5\textwidth, angle=0]{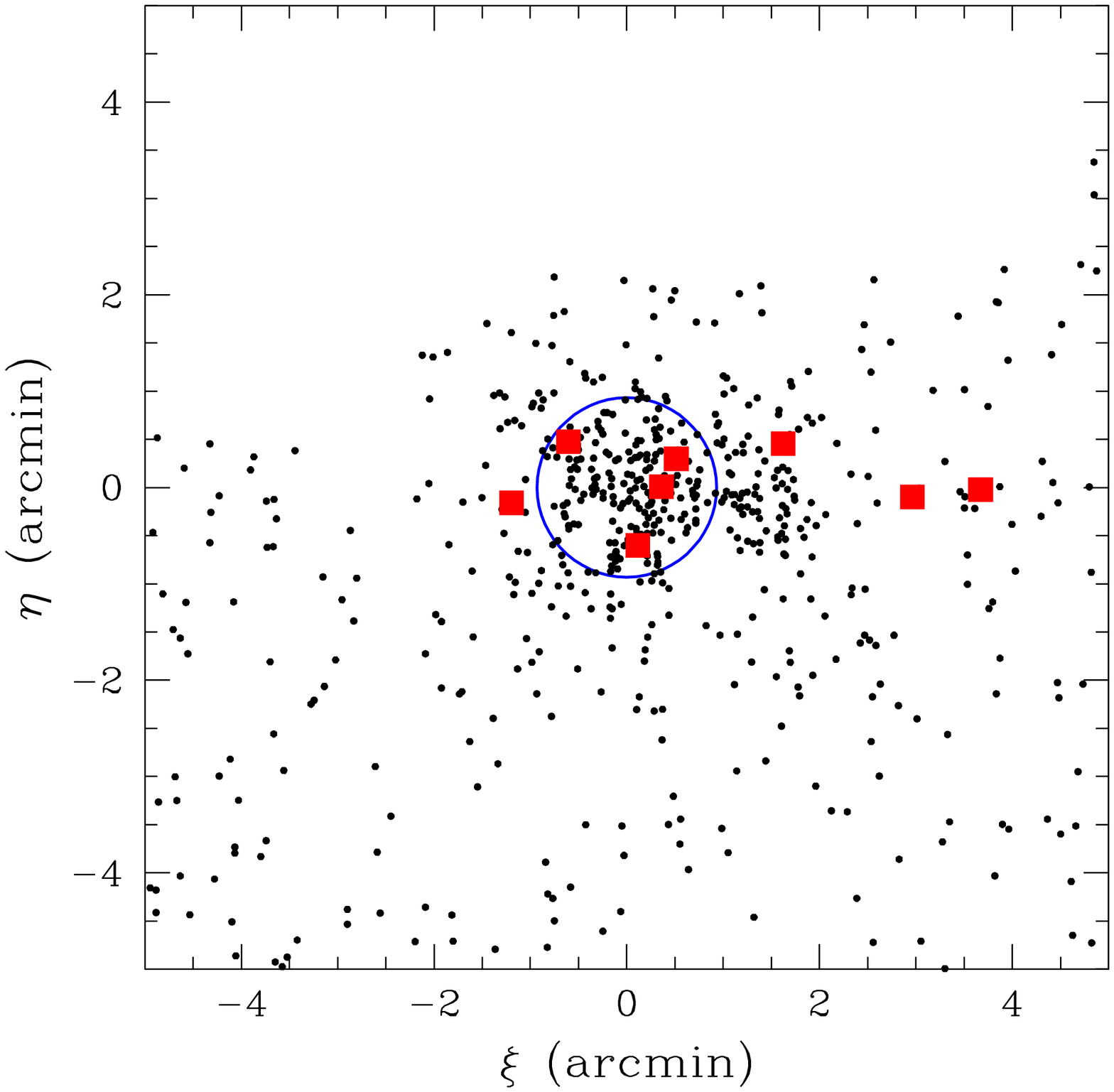}
\caption{Spatial distribution of stars in the regions of  \andxv\ (top) and \andxvi\ bottom
Candidate member stars  identified spectroscopically are highlighted as robust (filled squares)
likely (filled triangles), and in the case of \andxv\ three Galactic foreground stars which were previously identified as the tip of the RGB (pentagons).
Circular half-light radii are overlaid.
\label{fig:and15-vrad}}
\end{figure}


\section{ACKNOWLEDGMENTS}
The data presented herein were obtained at the W.M.
Keck Observatory, which is operated as a scientific partnership among
the California Institute of Technology, the University of California and
the National Aeronautics and Space Administration. The Observatory was
made possible by the generous financial support of the W.M. Keck
Foundation.

\begin{table*}
\begin{center}
\caption{Properties of candidate member stars in \andxv\ centered at $\alpha$ = 01h 14m 18.7s, $\delta$ = 38$^\circ$ 07$'$ 03$"$.}
\begin{tabular}{@{\extracolsep{-1.5pt}}llcccccc}
\noalign{\medskip}
\hline
$\alpha$ (J2000) & $\delta$ (J2000) & v$_r$ ($\kms$) & [F e/H]$_{spec}$ & S/N$_{\rm ctm}$\,$^b$ & [F e/H]$_{phot}$ & $g$-mag & $i$-mag \\
\hline
01:14:16.28& 38:05:49.6& -360.0$\pm$6.3& -2.4& 6.2& -1.45& 22.75& 21.26 \\
01:14:16.34& 38:06:08.2& -336.6$\pm$5.6& -2.0& 8.8& -1.47& 22.50& 20.86 \\
01:14:24.52& 38:06:14.6& -324.2$\pm$18.3& -0.2& 1.6& -1.59& 23.81& 22.78 \\
01:14:16.30& 38:06:25.9& -323.5$\pm$14.7& -2.5& 6.3& -1.61& 22.59& 21.19 \\
01:14:21.64& 38:06:31.0& -278.5$\pm$52.4& -0.0& {1.9} & -1.93& 23.54& 22.56 \\
01:14:23.84& 38:08:43.3& -335.0$\pm$23.4& -0.5& 2.9& -1.58& 23.44& 22.32 \\
01:14:17.94& 38:10:18.2& -334.6$\pm$6.1& -2.5& 6.7& -1.58& 22.64& 21.24 \\
\hline
01:14:21.90& 38:09:13.0& -296.6$\pm$12.5& -0.0& {1.0} & -1.30& 24.95& 23.97 \\
01:14:15.43& 38:05:54.2& -373.6$\pm$21.0& -1.2& 1.8& -1.61& 23.07& 21.97 \\
01:14:20.70& 38:05:58.9& -341.8$\pm$25.3& -1.5& 3.8& -1.46& 23.19& 21.93 \\
01:14:14.86& 38:06:19.8& -302.2$\pm$22.5& -1.6& 3.0& -1.73& 22.90& 21.81 \\
01:14:15.74& 38:06:51.8& -296.1$\pm$22.2& -1.5& 2.7& -1.85& 22.80& 21.71 \\
01:14:17.04& 38:07:23.8& -374.1$\pm$16.3& -1.4& 2.5& -1.44& 23.27& 22.06\\ 
\hline
\hline
\end{tabular}
\end{center} 
$^a$ Candidate member stars in And\,XV are divided into a robust group (top 7) and a more tentative group (bottom 6).\\
$^b$ Signal to noise in the continuum, estimated over the 250\AA\ region surrounding the CaT.
\end{table*}

\begin{table*}
\begin{center}
\caption{Properties of candidate member stars in \andxvi\ centered at $\alpha$ = 0h 59m 30.0s, $\delta$ = 32$^\circ$ 22$'$ 30$"$.}
\begin{tabular}{@{\extracolsep{-1.5pt}}llcccccc}
\noalign{\medskip}
\hline
$\alpha$ (J2000) & $\delta$ (J2000) & v$_r$ ($\kms$) & [F e/H]$_{spec}$ & S/N$_{\rm ctm}$ & [F e/H]$_{phot}$ & $g$-mag & $i$-mag \\
\hline
00:59:29.54 & 32:21:59.9& -373.5$\pm$23.1& -2.7 & 3.3 &-2.24& 22.38 &21.25\\ 
00:59:23.33 &32:22:26.4& -369.4$\pm$11.9 &-2.2 & 6.2 &-2.08 &22.07 &20.78 \\
00:59:43.04 &32:22:30.0& -380.1$\pm$17.8 &-1.2 & 1.1 &-2.48 &23.84 &22.99 \\
00:59:46.39 &32:22:34.5& -389.0$\pm$10.0 &-1.6 & 9.1 &-1.98 &21.85 &20.40 \\
00:59:30.71 &32:22:36.4& -409.8$\pm$16.9 &-2.3 & 6.8 &-2.23 &21.99 &20.74 \\
00:59:31.43 &32:22:53.8& -400.5$\pm$21.9 &-0.9 & 6.8 &-2.39 &21.98 &20.69 \\
00:59:36.69 &32:23:03.3& -391.3$\pm$24.8 &-1.8 & 2.6 &-2.31 &22.95 &21.94 \\
00:59:26.12 &32:23:04.5& -381.4$\pm$12.2 &-2.4 & 6.0 &-2.09 &22.12 &20.86 \\
\hline
\hline
\hline
\end{tabular}
\end{center} 
\end{table*}

\newcommand{\mnras}{MNRAS}
\newcommand{\pasa}{PASA}
\newcommand{\nat}{Nature}
\newcommand{\araa}{ARAA}
\newcommand{\aj}{AJ}
\newcommand{\apj}{ApJ}
\newcommand{\apjl}{ApJ}
\newcommand{\apjs}{ApJSupp}
\newcommand{\aap}{A\&A}
\newcommand{\aaps}{A\&ASupp}
\newcommand{\pasp}{PASP}

\begin{figure}
\centering
\includegraphics[width=0.5\textwidth, angle=0]{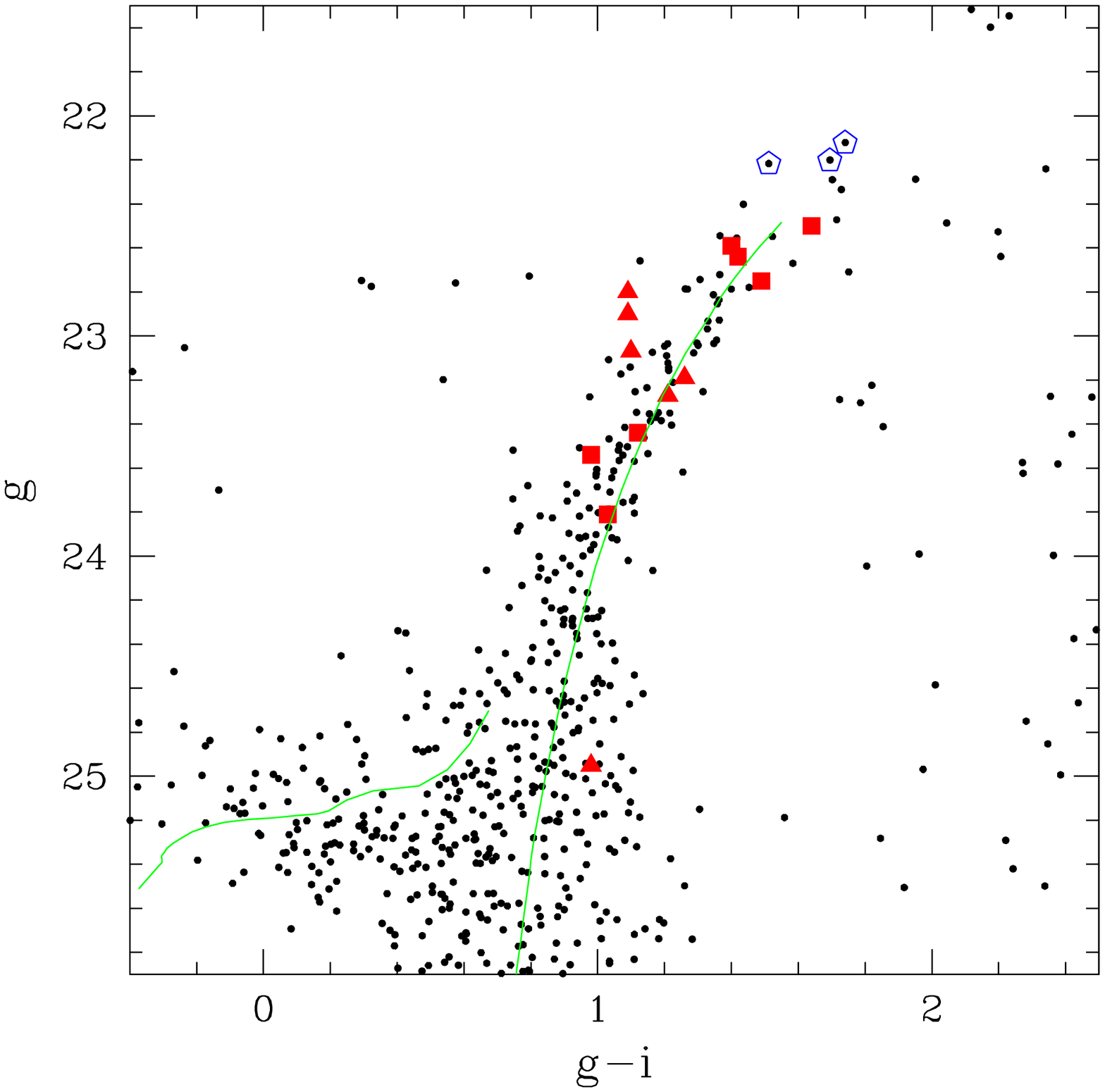}
\includegraphics[width=0.5\textwidth, angle=0]{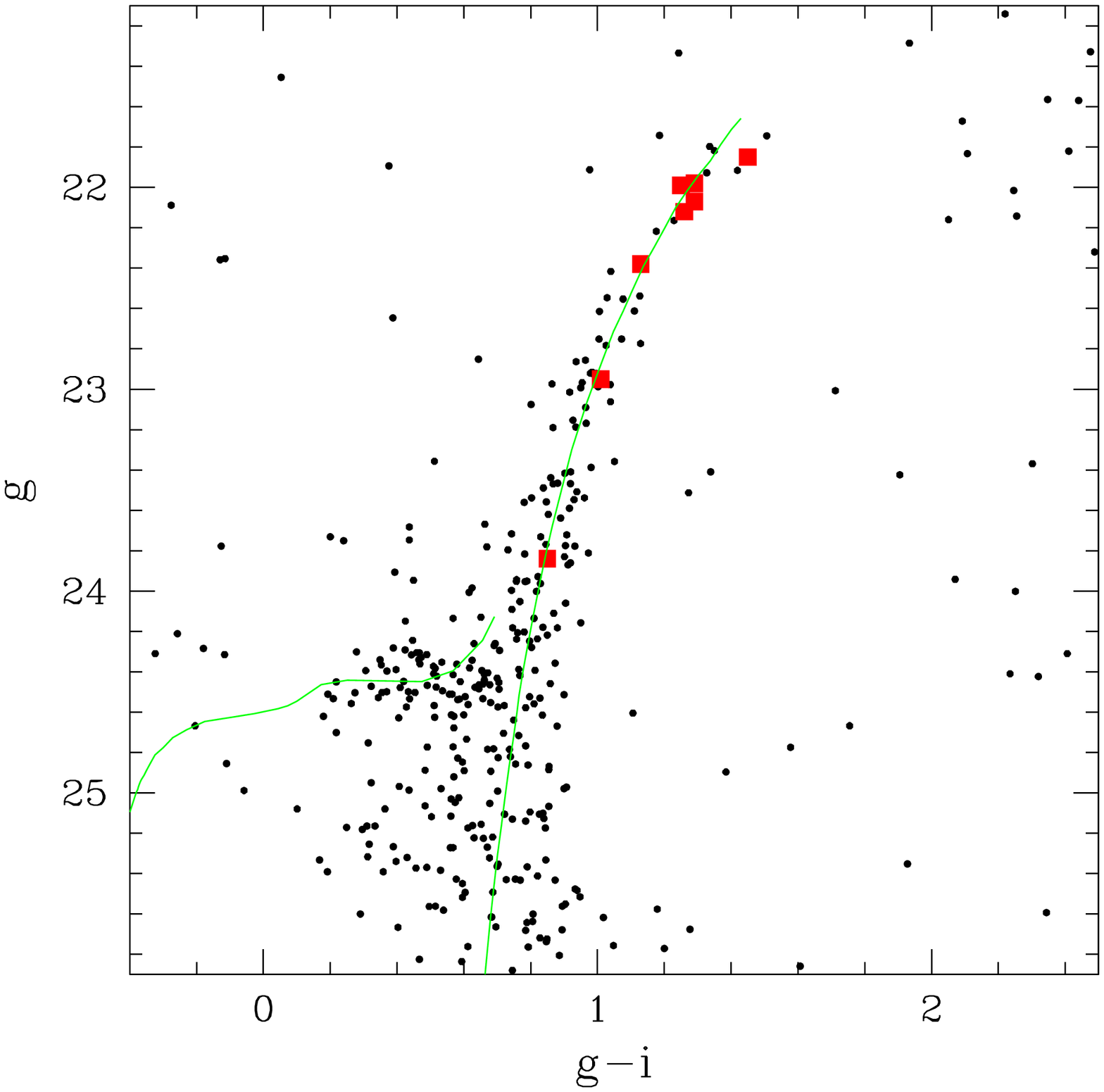}
\caption{CFHT-MegaCam colour-magnitude diagrams of stars 
within a three arcminute radius of And\,XV (top) and And\,XVI (bottom). The MegaCam mags are calibrated on
SDSS-like system as AB mags. The red giant branches of both dSphs are clearly visible, while the horizontal branch is reasonably detected in both cases. 
 Candidate member stars from the DEIMOS spectroscopy are 
highlighted (filled squares -- robust members,  filled triangles -- tentative members).
In the case of And\,XV, three stars 
previously assumed to lie near the TRGB have been shown by 
their velocities and Na{\sc I} doublet EWs to be Galactic foreground 
(open pentagons). The revised TRGB and horizontal branch distance to And\,XV is 770~kpc as described in the text. 
13~Gyr old Padova isochrones are overlaid, at the TRGB distance and median metallicity obtained from member stars, providing a reasonable fit in both cases.
\label{fig:and15-memb}}
\end{figure}

\begin{figure}
\centering
\includegraphics[width=0.5\textwidth, angle=0]{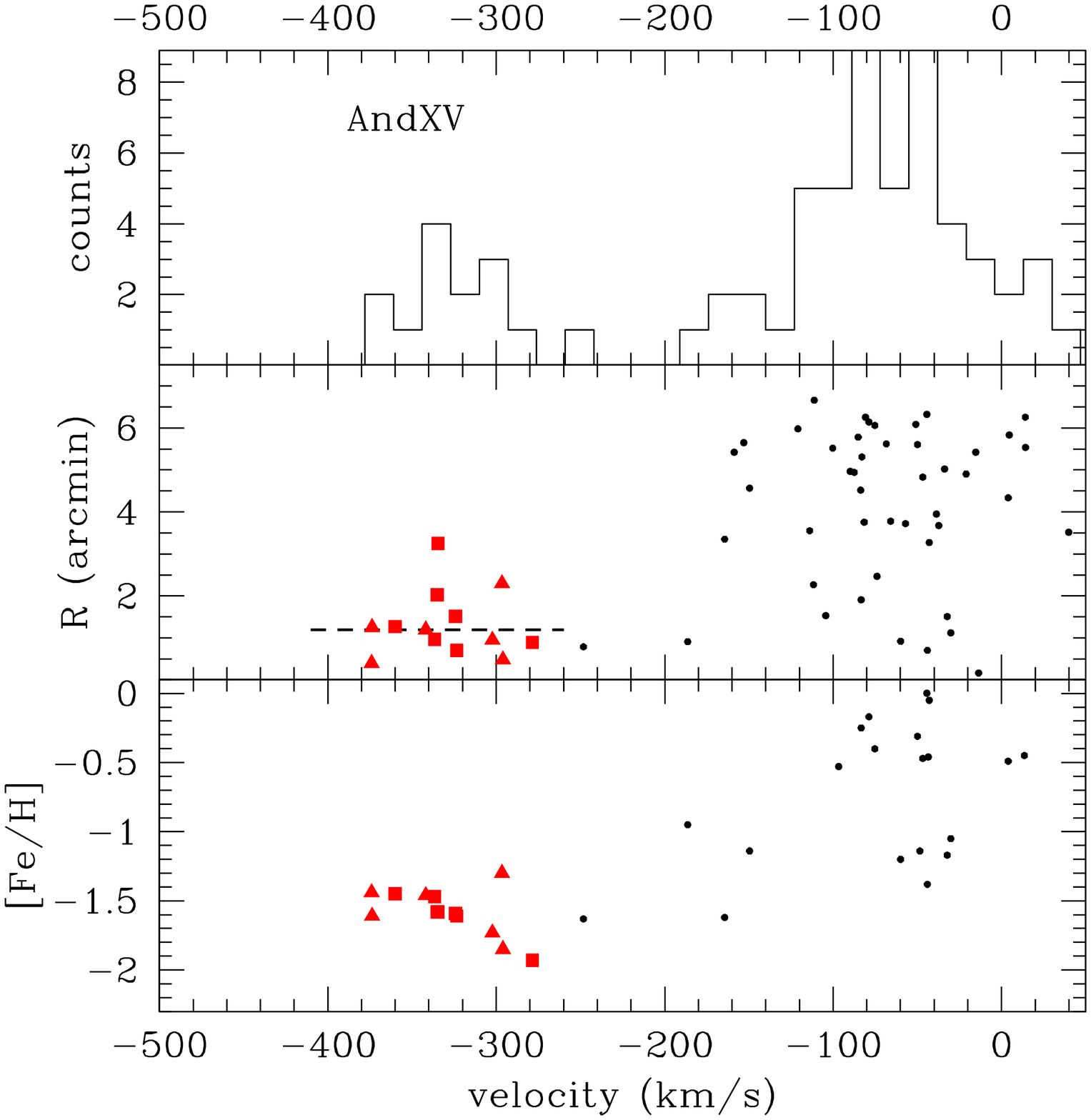} 
\includegraphics[width=0.5\textwidth, angle=0]{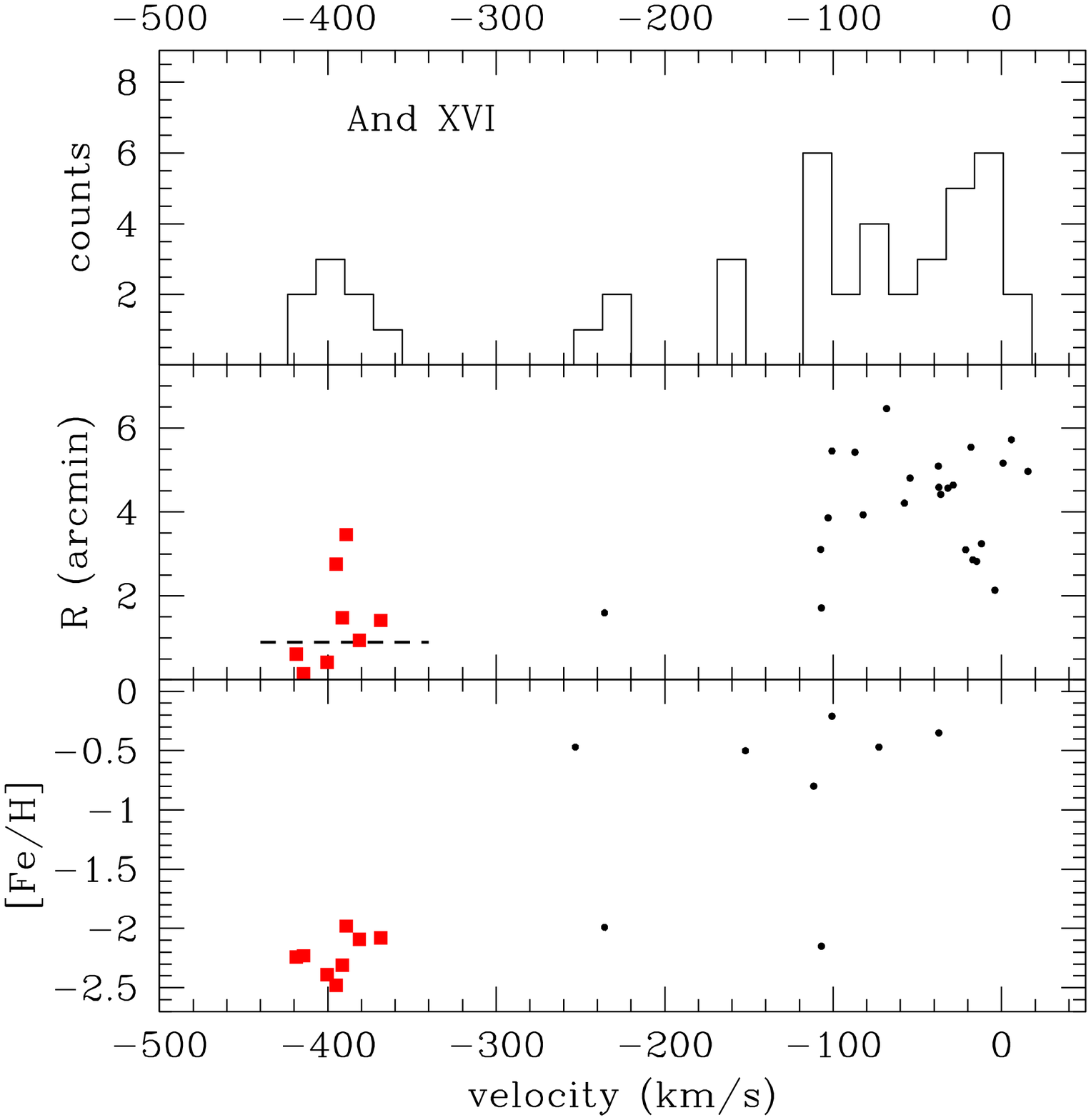} 
\caption{Distribution of stars in \andxv\ (top), \andxvi\ (bottom) as a function of their radial velocity
(upper panels). The stars are then shown as a function of radius 
from the centers of the dSphs, with half-light radius drawn as a dashed line (middle panels). Photometric 
[Fe/H] as described in the text is shown on the bottom panels, 
revealing the tight range in metallicities of \andxv\ ([Fe/H]$\sim$-1.6), 
\andxvi\ ([Fe/H]$\sim$-2.2).
Symbols are highlighted as in Figs~1\&2.
\label{fig:and16-vrad}}
\end{figure}

\begin{figure}
\centering
\includegraphics[width=0.5\textwidth, angle=0]{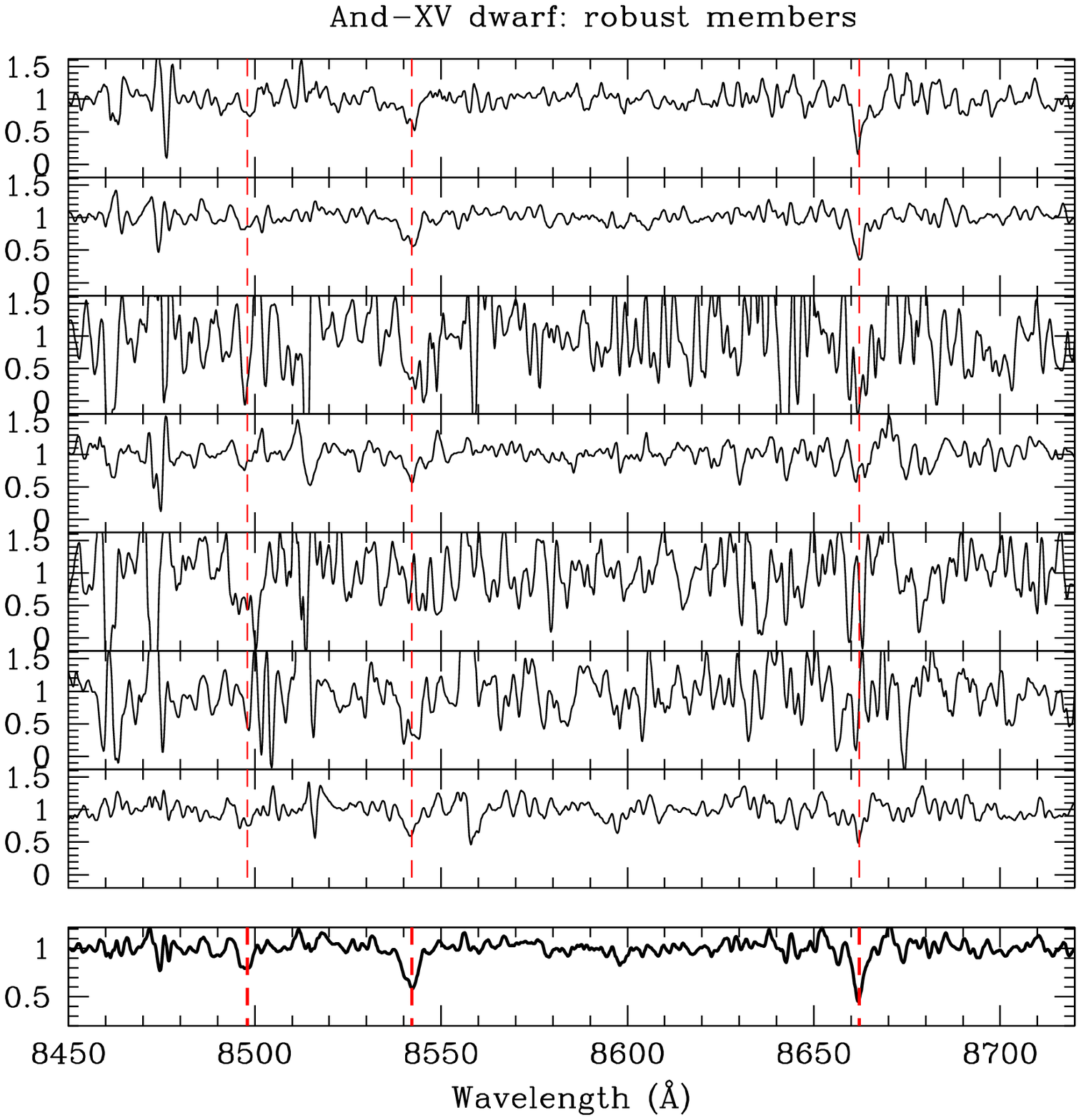}
\vskip-1cm
\includegraphics[width=0.5\textwidth, angle=0]{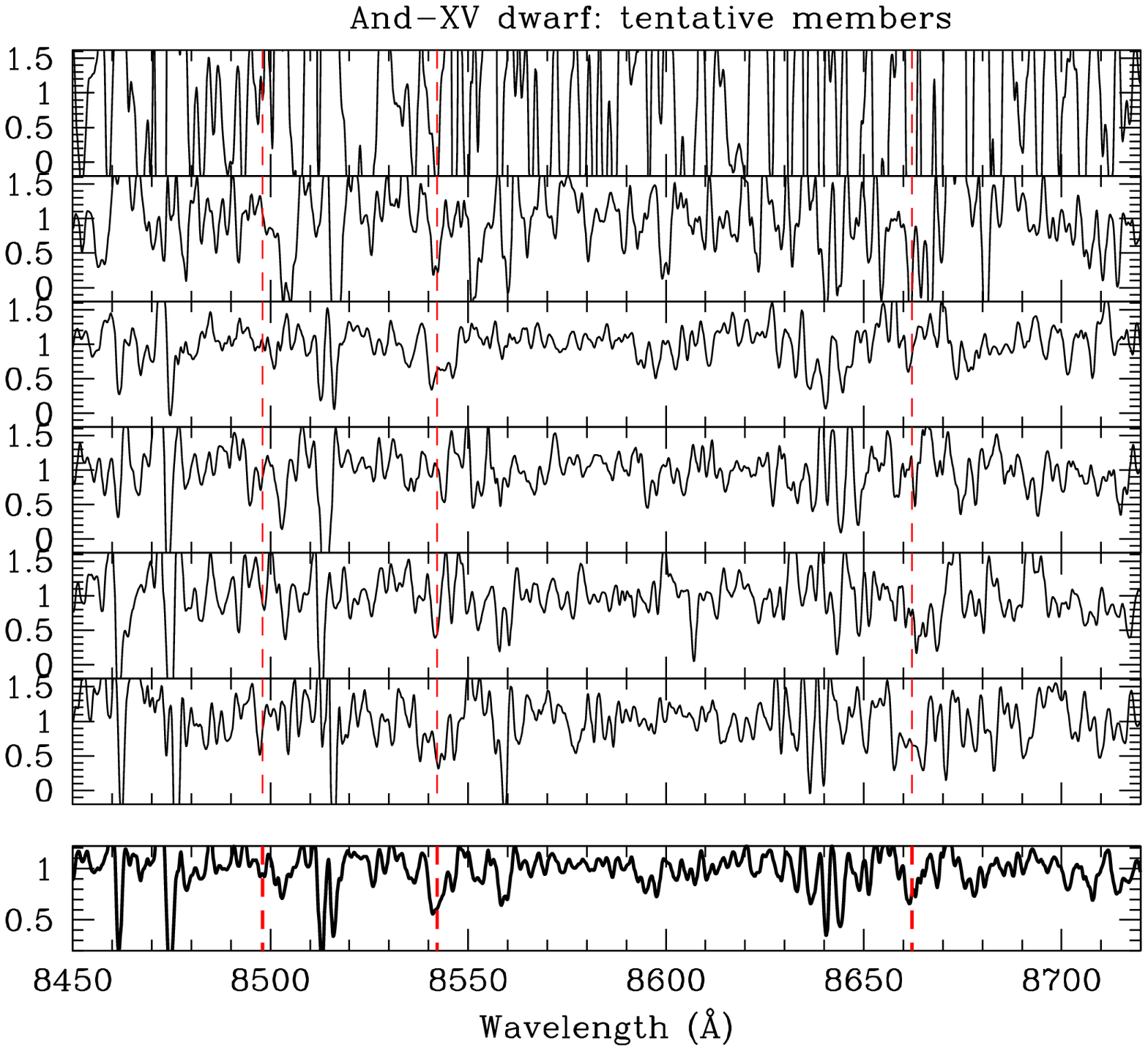}
\vskip-1.5cm
\caption{Spectra of candidate member stars 
in And\,XV. Top panel shows robust members with
cross correlation peak $> 0.2$. Bottom panel shows more tentative 
member stars with cross correlation peak $< 0.2$, 
but still lying on the well defined RGB of And\,XV. The inverse 
variance weighted summed spectrum is shown in the bottom offset 
panels for each subsample. 
\label{fig:and16-memb}}
\end{figure}

\begin{figure}
\centering
\includegraphics[width=0.5\textwidth, angle=0]{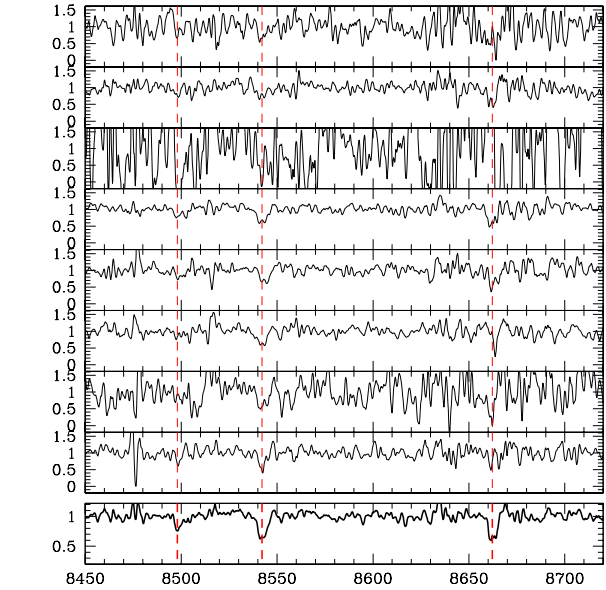}
\caption{Spectra of candidate member stars 
in And\,XVI. The inverse variance weighted summed spectrum is 
shown in the bottom offset panel.
\label{fig:and16-space-memb}}
\end{figure}

\begin{figure}
\centering
\includegraphics[width=0.5\textwidth, angle=-90]{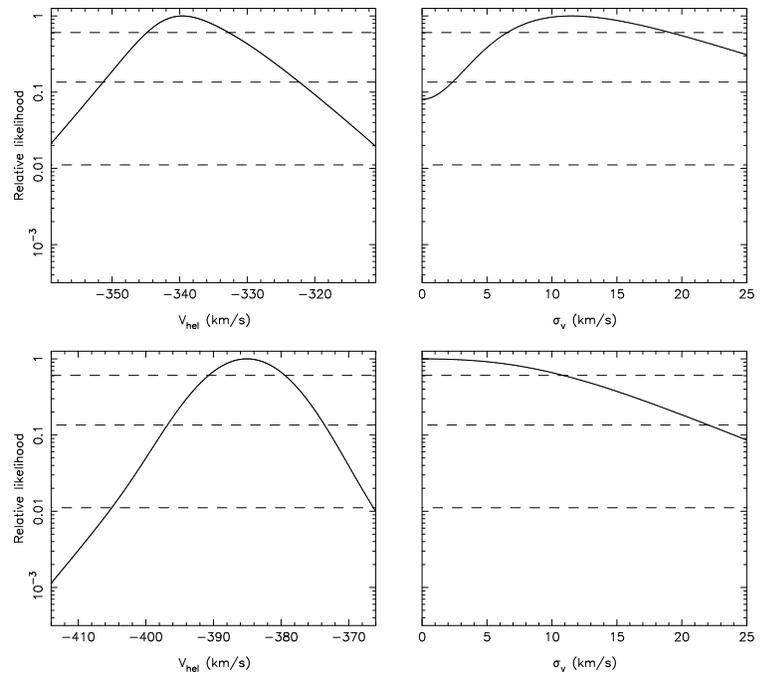}
\caption{Likelihood distributions of member stars in And\,XV 
(top) 
and And\,XVI (bottom). The point shows the most likely 
values of radial velocity and velocity dispersion, dashed lines 
showing the 1, 2, 3$\times \sigma$ distributions. For And\,XV, we show likelihood 
distributions of the seven best candidate members. 
The And\,XV dispersion is resolved 
with v$_r$ =-339$^{+7}_{-6} \kms$
and $\sigma_v = 11^{+7}_{-5} \kms$. 
The \andxvi\ dispersion is not quite resolved at $1\sigma$ with v$_r$ =-385$^{+5}_{-6} \kms$
and $\sigma=0^{+10}_{-indef} \kms$.
\label{fig:and16-space-memb}}
\end{figure}

\end{document}